\documentclass[aps,prb,amsmath,reprint,superscriptaddress]{revtex4-1}
\usepackage{amsmath,txfonts,bm,color,dcolumn,graphicx,physics,multirow}
\begin{document}

\title{On the accuracy of retinal protonated Schiff base models}
\author{Jae Woo Park}
\email{jwpk1201@northwestern.edu}
\author{Toru Shiozaki}
\affiliation{Department of Chemistry, Northwestern University, 2145 Sheridan Rd., Evanston, IL 60208, USA.}
\date{\today}

\begin{abstract}

We investigate the molecular geometries of the ground state and the minimal energy conical intersections (MECIs) between the ground and first excited states of the models for the retinal protonated Schiff base
in the gas phase
using the extended multistate complete active space second-order perturbation theory (XMS-CASPT2).
The biggest model in this work is the rhodopsin chromophore truncated between the $\epsilon$ and $\delta$ carbon atoms, which consists of 54 atoms and 12-orbital $\pi$ conjugation.
The results are compared with those obtained by the state-averaged complete active space self-consistent field (SA-CASSCF).
The XMS-CASPT2 results suggest that the minimum energy conical intersection associated with the so-called 13--14 isomerization
is thermally inaccessible, which is in contrast to the SA-CASSCF results.
The differences between the geometries of the conical intersections computed by SA-CASSCF and XMS-CASPT2 are ascribed to the fact that
the charge transfer states are more stabilized by dynamical electron correlation than the diradicaloid states.
The impact of the various choices of active spaces, basis sets, and state averaging schemes is also examined.

\end{abstract}

\maketitle

\section{Introduction}

Photoisomerization is among the non-adiabatic photochemical processes, which undergo electronic transitions near the conical intersections (CIs).\cite{Baerbook}
In particular, the photoisomerization of the retinal protonated Schiff base (RPSB) is one of the fastest processes in biology,
which occurs within several picoseconds.\cite{Schoenlein1991Science,Kandori1995JACS,Ernst2014CR,Bassolino2015JACS,Gozem2017CR}
The RPSB and related chromophores have been exhaustively studied using theoretical and computational methods.
For instance, geometry optimizations, potential energy scans,\cite{Cembran2004JACS,Andruniow2004PNAS,MunozLosa2011JCTC,Ben-Nun2002PNAS,Gozem2012JCTC,Mori2010JCP,Valsson2010JCTC,Gozem2012Science,Fantacci2004JPCA,Zhou2014JCC,Garavelli1998JACS,Gozem2014JCTC,Herbert2016ACR,Dokukina2017PhotochemPhotobiol,Sekharan2006BiophysJ,Valsson2012JPCL,Szymczak2008JCTC,Manathunga2017JPCL,Luk2015PNAS,Schapiro2011JACS,Manathunga2016JCTC,Gozem2017CR}
and nonadiabatic molecular dynamics (MD) simulations\cite{Frutos2007PNAS,Liu2016JPCB,Szymczak2008JCTC,Ruckenbauer2010JPCA,Li2011JCTC,Schapiro2011JACS,Manathunga2016JCTC,Manathunga2017JPCL,Valentini2017ACIE,Virshup2009JPCB,Luk2015PNAS,Gozem2017CR}
of truncated model chromophores have been performed in such studies.
Because the photoisomerization of RPSB involves transitions between the $\mathrm{S}_1$ and $\mathrm{S}_0$ electronic states,
multireference wave function methods that provide a balanced description of the ground and excited states have often been used to describe the electronic structure of the model chromophores.
They include, for instance, state-averaged complete active space self-consistent field (SA-CASSCF),\cite{Garavelli1998JACS,Ben-Nun2002PNAS,Fantacci2004JPCA,Cembran2004JACS,Andruniow2004PNAS,Frutos2007PNAS,Szymczak2008JCTC,Mori2010JCP,Valsson2010JCTC,Ruckenbauer2010JPCA,Li2011JCTC,MunozLosa2011JCTC,Schapiro2011JACS,Gozem2012Science,Gozem2012JCTC,Valsson2012JPCL,Liu2016JPCB,Manathunga2016JCTC,Manathunga2017JPCL,Valentini2017ACIE,Luk2015PNAS,Gozem2017CR}
complete active space second-order perturbation theory (CASPT2),\cite{Fantacci2004JPCA,Cembran2004JACS,Andruniow2004PNAS,Frutos2007PNAS,Szymczak2008JCTC,Mori2010JCP,Valsson2010JCTC,Schapiro2011JACS,MunozLosa2011JCTC,Valsson2012JPCL,Gozem2012Science,Gozem2012JCTC,Zhou2014JCC,Liu2016JPCB,Manathunga2016JCTC,Manathunga2017JPCL,Dokukina2017PhotochemPhotobiol,Luk2015PNAS,Sekharan2006BiophysJ,Gozem2017CR}
and multireference configuration interactions (MRCI).\cite{Szymczak2008JCTC,Ruckenbauer2010JPCA,Gozem2012JCTC,Gozem2017CR}

The optimization of the molecular geometries of the small RPSB models using CASPT2 was first performed a few decades ago, in which the equilibrium geometries for two protonated Schiff bases ($\mathrm{C}_5\mathrm{H}_6\mathrm{NH}_2^+$ and $\mathrm{C}_{10}\mathrm{H}_{12}\mathrm{NH}_2^+$) and the conical intersection of $\mathrm{C}_5\mathrm{H}_6\mathrm{NH}_2^+$ were considered.\cite{Page2003JCC}
The ground-state equilibrium structures of the protonated Schiff bases up to $\mathrm{C}_9\mathrm{H}_{10}\mathrm{NH}_2^+$ (PSB5) have recently been studied using MS-CASPT2.\cite{Valsson2010JCTC,Dokukina2017PhotochemPhotobiol}
The nuclear gradients were calculated using finite difference formulas in these studies, and the derivative coupling elements were computed using SA-CASSCF.\cite{Page2003JCC,Valsson2010JCTC,Dokukina2017PhotochemPhotobiol}
The use of numerical gradients, however, does limit the size of the systems to be investigated, because the number of the nuclear degrees of freedom ($3N$) is multiplied to the computational expense of the underlying quantum chemical methods.

Though several studies\cite{Mori2010JCP,Liu2016JPCB} (using a partially contracted variant, or RS2\cite{Celani2003JCP,Mori2009CPL,Shiozaki2011JCP3}) have demonstrated for small molecules
the importance of including dynamical correlation in photodynamics simulations,  
such simulations for large RPSB models had not been possible until recently, due to the lack of analytical nuclear gradient code for CASPT2 with full internal contraction.
Therefore, many of the previous studies have used a hybrid method in which geometries are obtained using CASSCF and energies are computed using CASPT2 (often denoted as CASPT2//CASSCF).\cite{Fantacci2004JPCA,Cembran2004JACS,Andruniow2004PNAS,Frutos2007PNAS,Schapiro2011JACS,MunozLosa2011JCTC,Gozem2012Science,Gozem2012JCTC,Zhou2014JCC,Manathunga2016JCTC}
For the protonated Schiff bases, the CASPT2 energies at the CASPT2 and CASSCF optimized geometries of the excited states relative to the $\mathrm{S}_0$ minimum have shown to differ by 0.11--0.28 eV,\cite{Page2003JCC}
and the CASPT2 energy gaps at the conical intersection geometries computed by CASSCF have been around 0.2 eV.\cite{Gozem2012JCTC}
Very recently, our group has addressed this problem and reported an efficient parallel implementation of the analytical nuclear gradients\cite{MacLeod2015JCP,Vlaisavljevich2016JCTC,Park2017JCTC2}   and derivative couplings\cite{Park2017JCTC} for CASPT2 and its multistate variants.
This new development has motivated us to revisit the truncated RPSB models for retinal using CASPT2.

In this work, we have optimized the molecular structures of the RPSB model chromophores using the extended multistate CASPT2 (XMS-CASPT2).
The biggest model consists of 54 atoms and $\pi$ conjugation with 12 orbitals.
We have shown how the energies change as a function of the conjugation lengths in these model chromophores
and have examined the impact of various basis sets, active spaces, and state averaging schemes. 
The physical origin of such changes is discussed.

\section{Model Chromophores}

\begin{figure}
	\includegraphics[width=0.90\linewidth]{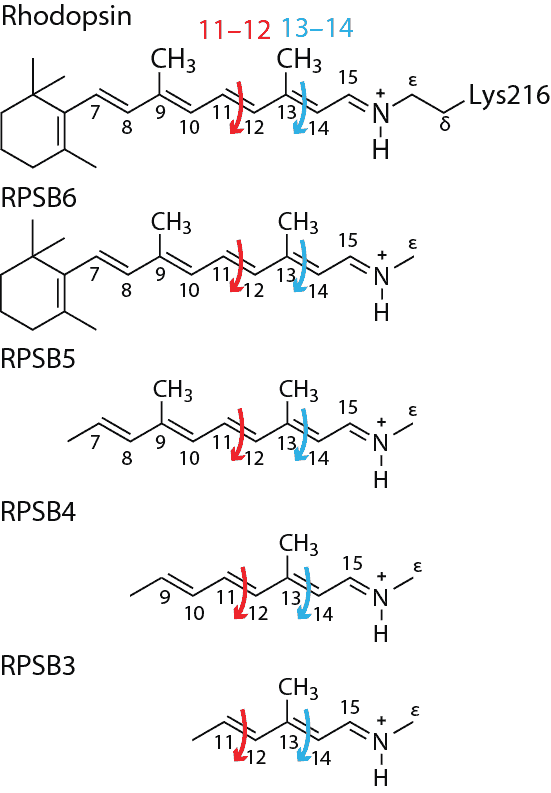}
	\caption{The structure of rhodopsin and model chromophores used in this study. The main photoisomerization coordinates are highlighted with red (11--12) and blue (13--14) arrows.}
	\label{figmodels}
\end{figure}

Four truncated models of the G protein coupled receptor rhodopsin are studied (see Fig.~\ref{figmodels} for their structures and atom labels).
The largest model, RPSB6, is a rhodopsin chromophore with a truncated single bond between the $\epsilon$ and $\delta$ carbons.
This RPSB6 model includes all the double bonds in the chromophore,
and is used as a reference in this study.
The other three models have an additional truncated single bond on the side of cyclohexene ring.
These models are designed to have different numbers of double bonds in the carbon chain,
while the $\mathrm{C}_{15}$--N double bond is included in all models.
For brevity, we will denote these model chromophores as RPSB$n$ in the following discussions,
where $n$ is the number of double bonds in these chromophores.

These chromophores can be regarded as modifications of the models that have been investigated in previous studies.
Compared to the minimal model for the RPSB photoisomerization [PSB3 ($\mathrm{C}_5\mathrm{H}_6\mathrm{NH}_2^+$)],\cite{Page2003JCC,Szymczak2008JCTC,Mori2010JCP,Ruckenbauer2010JPCA,Gozem2012JCTC,Gozem2014JCTC,Liu2016JPCB,Zhou2014JCC,Szymczak2008JCTC,Valsson2010JCTC,Fantacci2004JPCA,Gozem2012Science}
RPSB3 has two additional methyl groups at the ends of the chromophore and a branched methyl group attached to the $\mathrm{C}_{13}$ atom.
RPSB3 corresponds to the model in the recent studies on the QM/MM photodynamics\cite{Manathunga2016JCTC} and optomechanical control of photoisomerization.\cite{Valentini2017ACIE}
RPSB4 without the branch and edge methyl groups corresponds to the model chromophore in Ref.~\onlinecite{Cembran2004JACS}.
RPSB5 is a similar model to that studied in Refs.~\onlinecite{Ben-Nun2002PNAS} and \onlinecite{Herbert2016ACR},
but with a methyl group on the C-terminal.
RPSB5 without the branch or edge methyl groups was also considered in some studies.\cite{MunozLosa2011JCTC,Garavelli1998JACS}
RPSB6 corresponds to the model chromophore in the previous CASSCF or DFT investigations,
although some of them used the model without the methyl group attached to the nitrogen atom.\cite{Virshup2009JPCB,Schapiro2011JACS,Zhou2014JCC,Gozem2012Science,Valsson2012JPCL,Walczak2015PCCP,Manathunga2017JPCL,Ping2017CPL}
RPSB6 is also the model chromophore in several QM/MM studies.\cite{Andruniow2004PNAS,Frutos2007PNAS,Luk2015PNAS}
RPSB6 model truncated at the edge of the double bond on the cyclohexene ring was also employed in a QM/MM study.\cite{Li2011JCTC}
The minimum energy structures and pathways in RPSB3, RPSB4, RPSB5 without the edge and branch methyl groups were also optimized using CASSCF and quantum Monte Carlo.\cite{Valsson2010JCTC}

\begin{figure*}[tb]
	\includegraphics[width=0.90\linewidth]{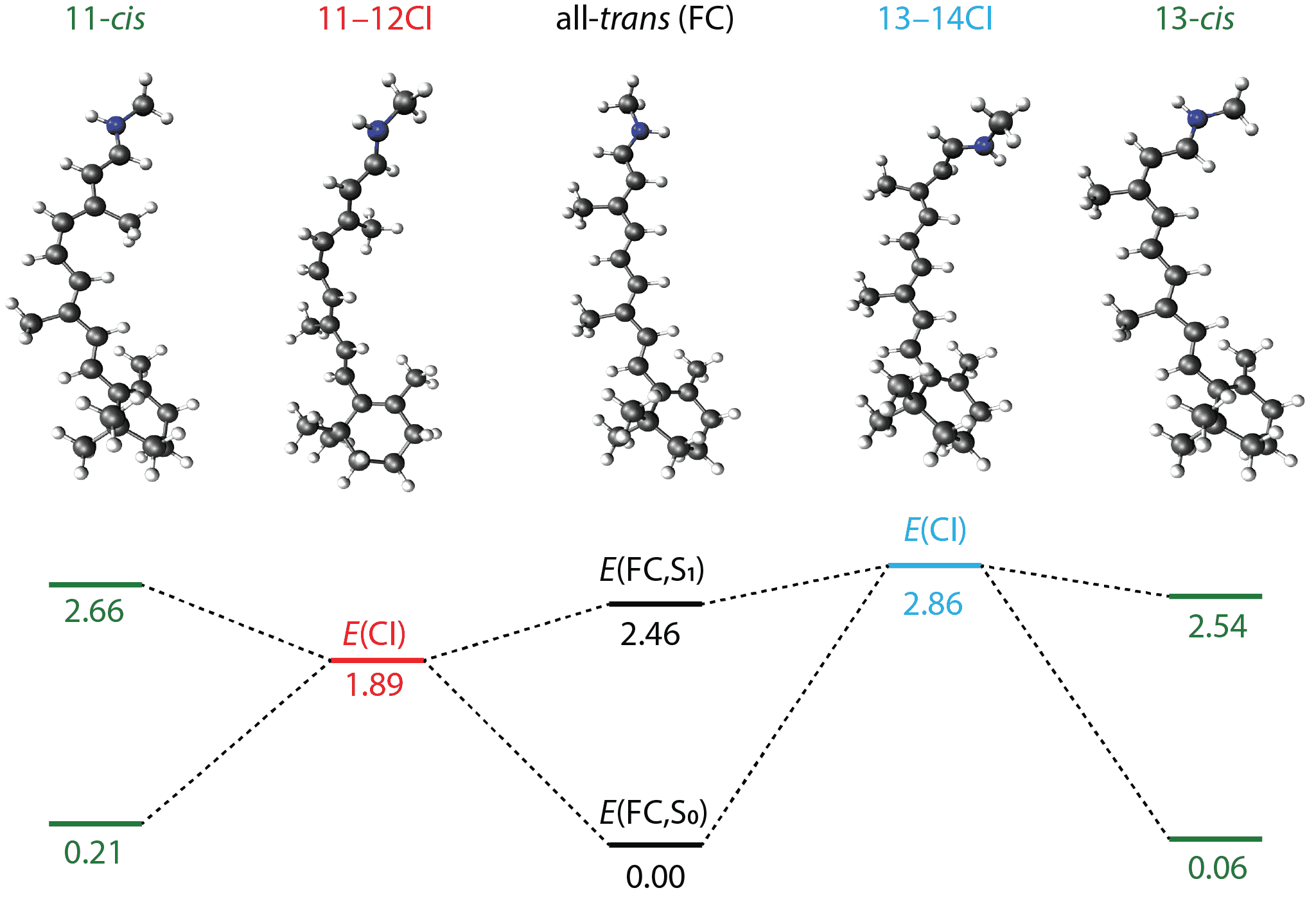}
	\caption{CASPT2 potential energy landscape of RPSB6 associated with the 11--12 and 13--14 photoisomerization. The energies are reported in eV with respect to the $\mathrm{S}_0$ energy at the all-\textit{trans} geometry. The (12$e$,12$o$) active space consisting of all of the $\pi$ orbitals was used. Three states were included in the calculations, one of which is not shown.
	\label{figglobal}}
\end{figure*}

\section{Computational Details}
The ground-state equilibrium geometries at the all-\textit{trans}, 11-\textit{cis}, and 13-\textit{cis} geometries, and the MECIs that correspond to the 11--12 and 13--14 photoisomerizations were optimized using the SA-CASSCF and XMS-CASPT2 methods.
These MECIs will be denoted as 11--12CI and 13--14CI in the following.

In the SA-CASSCF and XMS-CASPT2 calculations, the states to be included should be specified.
At the Franck--Condon (FC) point, the three lowest states of the RPSB are labeled as $1\mathrm{A}_g$, $1\mathrm{B}_u$, and $2\mathrm{A}_g$ states,
following the notation for polyenes in the $C_{2h}$ symmetry.\cite{Gozem2012JCTC,Gozem2012Science,Manathunga2017JPCL,Gozem2017CR}
The $1\mathrm{A}_g$ state corresponds to the covalent state at the FC point and the diradicaloid state at the CIs.
The $1\mathrm{B}_u$ state is the charge transfer state,
and the $2\mathrm{A}_g$ state is another diradicaloid state with two or more diradical sites.\cite{Gozem2012JCTC,Gozem2012Science,Manathunga2017JPCL,Gozem2017CR}
The main results were obtained by including the three lowest singlet states in the SA-CASSCF and XMS-CASPT2 calculations,
while we also present in Sec.~\ref{SA2sec} the results with a scheme that includes the two lowest singlet states.

Active spaces were the full valence $\pi$-spaces [(6\textit{e},6\textit{o}), (8\textit{e},8\textit{o}), (10\textit{e},10\textit{o}) and (12\textit{e},12\textit{o}) for RPSB3, RPSB4, RPSB5 and RPSB6, respectively].
Smaller active spaces [(8\textit{e},8\textit{o}) and (10\textit{e},10\textit{o})] were also tested for RPSB6 in Sec.~\ref{activesec}.
In the XMS-CASPT2 calculations, a vertical shift of $0.5 E_\mathrm{h}$ was used.\cite{Roos1995CPL}
The so-called SS-SR contraction scheme was employed.\cite{Vlaisavljevich2016JCTC}
The calculations were performed using the cc-pVDZ basis set and its corresponding JKFIT basis set for density fitting unless otherwise specified.
The MECIs were optimized using the gradient projection method.\cite{Bearpark1994CPL}
All quantum chemistry calculations were performed using the \textsc{bagel} program package.\cite{bagel,bagelreview}

\section{Results and Discussions}
Figure~\ref{figglobal} shows the XMS-CASPT2 potential energy landscape of RPSB6 associated with the isomerization dynamics.
In the following, we will discuss the accuracy of the truncated models and the computational approaches
(active spaces, basis sets, and state averaging schemes)
using the labels shown in Fig.~\ref{figglobal}.

\subsection{Truncated Models}

\begin{figure}
	\includegraphics[width=0.95\linewidth]{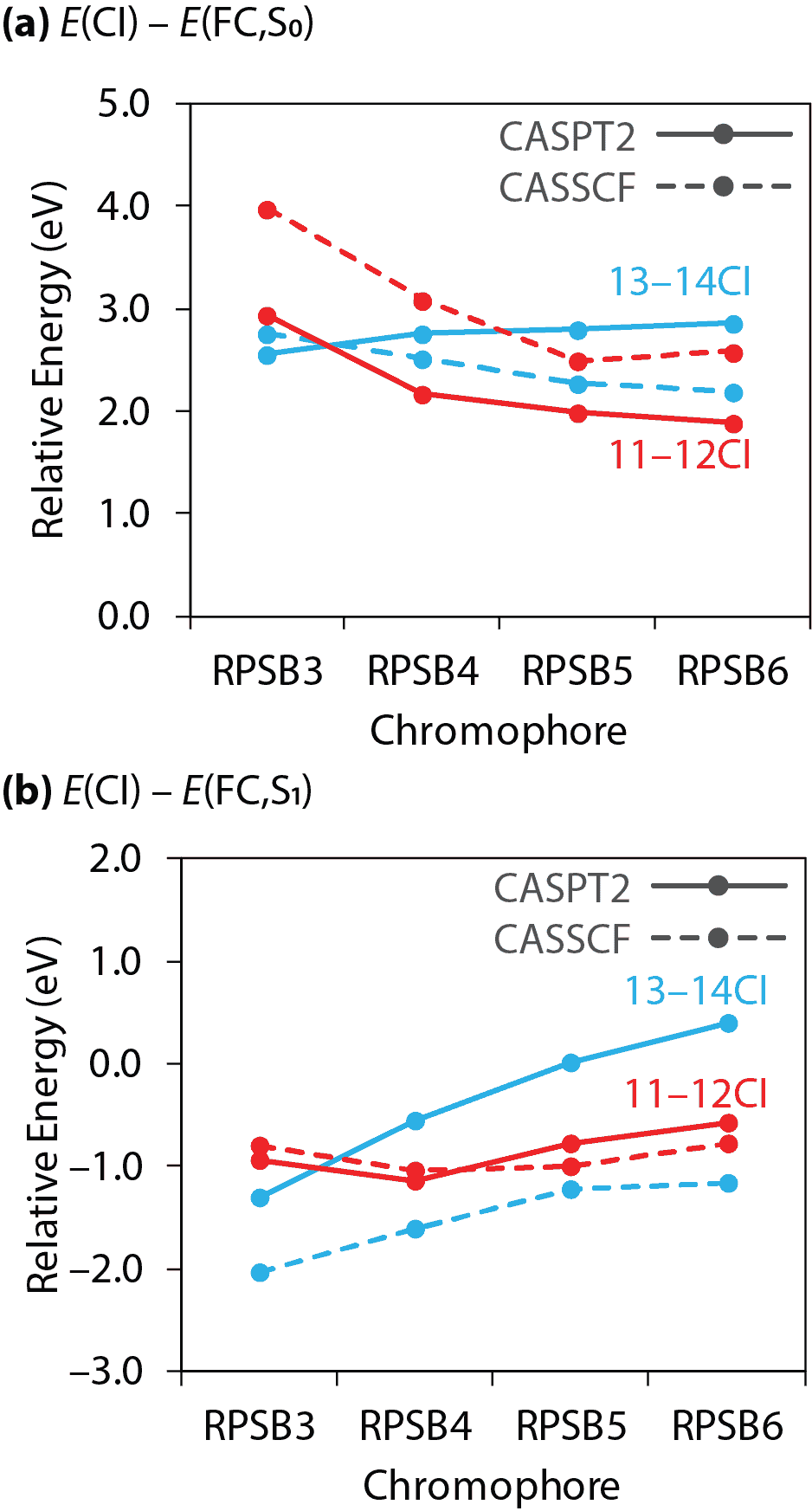}
	\caption{Trends of the energies (in eV) with respect to the chromophore length. The energies at the MECIs with respect to the length of the chromophore compared to (a) $\mathrm{S}_0$ and (b) $\mathrm{S}_1$ energies at the all-\textit{trans} FC point.}
	\label{figtrend}
\end{figure}

Figure~\ref{figtrend} shows the trends of the energies with respect to the length of the models.
The Cartesian coordinates of the optimized geometries are compiled in the supplemental online material.
As the $\pi$ conjugation of the chromophore becomes longer, the energies at the 11--12CI and 13--14CI become lower and higher, respectively, when computed using XMS-CASPT2 [Fig.~\ref{figtrend}(a)].
For RPSB5 and RPSB6, the energy at the 13--14CI is higher than the $\mathrm{S}_1$ energy at the all-\textit{trans} geometry [Fig.~\ref{figtrend}(b)].
This suggests that the 13--14CI is not thermally reachable from the all-\textit{trans} geometry.
This is consistent with the recent gas phase photoisomerization experiment, in which the 11-\textit{cis} and 9-\textit{cis} photoisomers were the major products and the 13-\textit{cis} photoisomer was the minor product.\cite{Coghlan2015PCCP}
The trend of the XMS-CASPT2 energies at the 13--14CI is not qualitatively reproduced by SA-CASSCF.

\begin{figure}
	\includegraphics[width=0.95\linewidth]{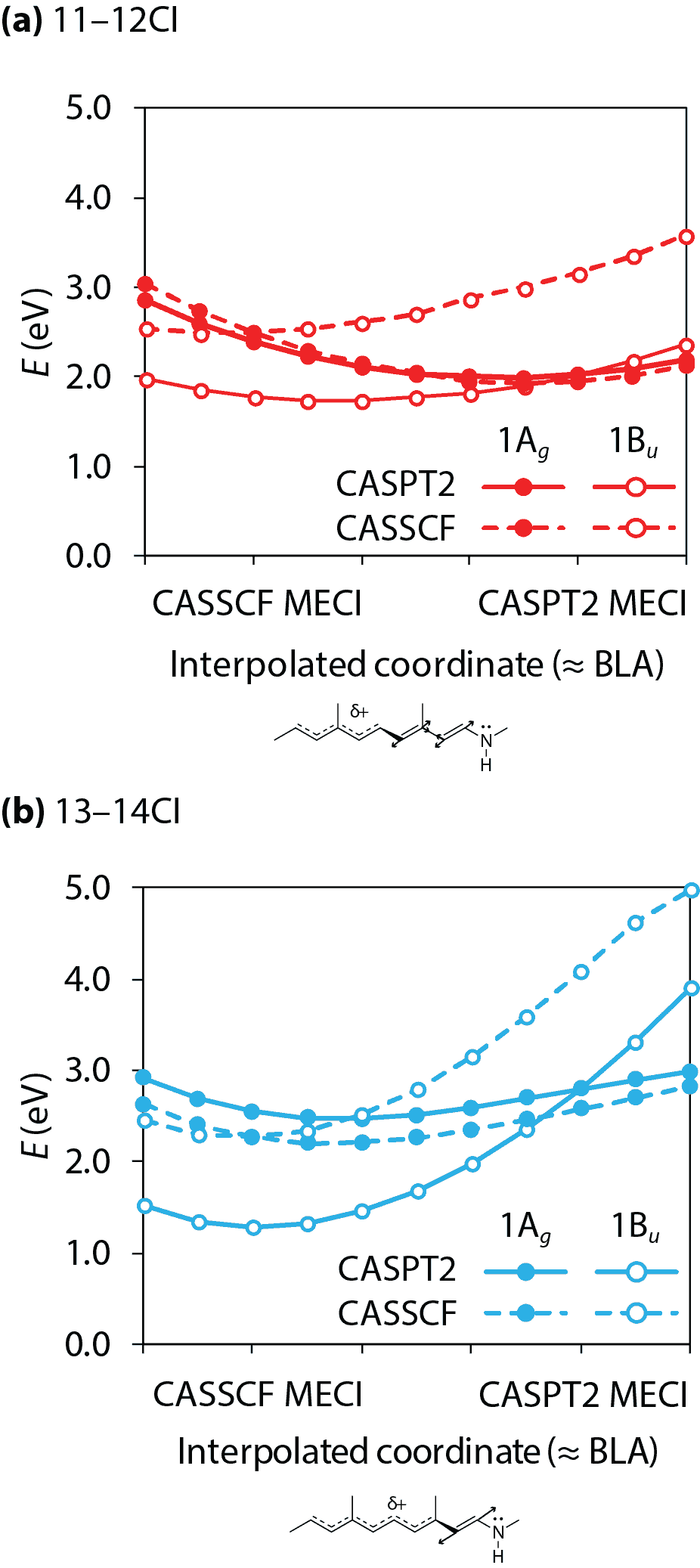}
	\caption{Energies (in eV) of the $\mathrm{S}_0$ and $\mathrm{S}_1$ states computed by XMS-CASPT2 (full) and SA-CASSCF (dotted) along the interpolated coordinates between the XMS-CASPT2 and SA-CASSCF MECIs in RPSB5.}
	\label{figinterpol}
\end{figure}

We scanned the XMS-CASPT2 and SA-CASSCF energies for RPSB5 along the interpolated coordinates connecting the MECIs calculated using XMS-CASPT2 and SA-CASSCF (Fig.~\ref{figinterpol}).
The diabatic states that cross at the MECI can be assigned to the charge transfer ($1\mathrm{B}_u$) and diradicaloid ($1\mathrm{A}_g$) states.
It is found that the $1\mathrm{B}_u$ state is more stabilized by dynamical correlation than the $1\mathrm{A}_g$ state at both MECIs;
the PT2 contributions to the energies for the $1\mathrm{B}_u$ state is 0.61~eV~(11--12CI) and 1.27~eV~(13--14CI) larger than those for the $1\mathrm{A}_g$ state at the SA-CASSCF MECI geometries of RPSB5.
This differential dynamical correlation effect is found to result in large structural changes at the MECIs.

The key structural differences between the MECIs obtained by SA-CASSCF and XMS-CASPT2 are the bond lengths.
With the PT2 corrections, the $\mathrm{C}_{12}$--$\mathrm{C}_{13}$ and $\mathrm{C}_{14}$--$\mathrm{C}_{15}$ bonds at the 11--12CI geometry elongate by 0.05~\AA, whereas
the $\mathrm{C}_{13}$--$\mathrm{C}_{14}$ bond contracts by 0.05~\AA.
These bond length changes correspond to the so-called bond-length alteration (BLA) coordinate,
which is a coordinate defined by the elongation of the double bonds and the contraction of the single bonds in the $1\mathrm{B}_u$ state (see also Fig.~\ref{figinterpol}).\cite{Gozem2012Science,Gozem2017CR}
Displacing the molecular geometry toward the FC point along the BLA coordinate stabilizes the diradicaloid diabatic state, as it alters the single and double bonds in the $1\mathrm{B}_u$ state.\cite{Gozem2012Science,Gozem2017CR}

The $\mathrm{C}_{14}$--$\mathrm{C}_{15}$ bond of the 13--14CI geometry elongates by 0.10~\AA~with the PT2 corrections,
which corresponds to the BLA coordinate for the 13--14CI.\cite{Gozem2012JCTC}
This is in agreement with the previous results for PSB3, where the bond length at the 13--14CI geometry was found to be 0.12~\AA~longer when computed with RS2 than the geometry with CASSCF.\cite{Liu2016JPCB}
The $\mathrm{C}_{12}$--$\mathrm{C}_{13}$--$\mathrm{C}_{14}$--$\mathrm{C}_{15}$ torsional angle computed by XMS-CASPT2 is about $15^\circ$ larger than the angle calculated by SA-CASSCF.
This change also leads to additional stabilization of the $1\mathrm{A}_g$ state relative to the $1\mathrm{B}_u$ state.
The shift of the 13--14CI toward the FC point along the BLA coordinate with dynamical correlation was also demonstrated for PSB3 using a variety of the quantum chemical methods [SI-SA-REKS, EOM-SF-CCSD(dT), MRCISD, CASPT2, XMSQDPT2, QD-NEVPT2].\cite{Gozem2012JCTC,Gozem2012Science,Gozem2014JCTC,Gozem2017CR}
Note that we also observed the same differential dynamical correlation effects in smaller models (RPSB3 and RPSB4); see the supplemental online material for details.

\begin{table}[t]
\caption{
Comparison between XMS-CASPT2 and XMS-CASPT2//SA-CASSCF. The vertical excitation energies at the FC point and the MECI energies relative to the respective $\mathrm{S}_0$ minimum are compared (in eV).}
\label{tabcaspt2casscf}
\begin{ruledtabular}
	\begin{tabular}{cddddd}
		& \multicolumn{1}{c}{FC} & \multicolumn{2}{c}{11--12CI\footnotemark[1]}  & \multicolumn{2}{c}{13--14CI\footnotemark[1]}  \\ \hline
        RPSB3 & 0.12     & 0.22 & 0.49 & -0.79 & 0.26\\
		RPSB4 & 0.11     & 0.09 & 0.52 & -1.32 & -0.14 \\
		RPSB5 & 0.11     & -0.33 & 0.28 & -1.63 & -0.36\\
		RPSB6 & 0.28     & -0.49 & -0.13 &  -1.82 & -0.73
	\end{tabular}
\end{ruledtabular}
\footnotetext[1]{ The difference between the two numbers corresponds to the energy gap in the CASPT2//CASSCF approach.} 
\end{table}

To quantitatively assess the accuracy of the XMS-CASPT2//SA-CASSCF protocol,
we calculated the XMS-CASPT2 energies at the SA-CASSCF optimized geometries for all the models (Table~\ref{tabcaspt2casscf}).
The $\mathrm{S}_0$ to $\mathrm{S}_1$ vertical excitation energy differences between XMS-CASPT2 and XMS-CASPT2//SA-CASSCF at the FC point are relatively small and are in the range of 0.11--0.28 eV.
The CI energy differences between the XMS-CASPT2 and XMS-CASPT2//SA-CASSCF energies are in the range of $-1.82$--0.52 eV, and are significantly larger than the differences at the FC point.
For RPSB6, $\mathrm{S}_0$ and $\mathrm{S}_1$ energies at both MECIs calculated using XMS-CASPT2//SA-CASSCF are below the energies computed using XMS-CASPT2.
The XMS-CASPT2 gap energies evaluated at the CASSCF optimized MECI geometries are in the range of 0.2--0.6 eV (11--12CI) and 1.0--1.4 eV (13--14CI).

\subsection{Active Spaces\label{activesec}}

\begin{table}[t]
	\caption{
		Dependence on the size of active spaces for RPSB6. The $\mathrm{S}_1$ energies at the FC point and MECIs are shown in eV relative to the $\mathrm{S}_0$ energy at the FC point.
        The MECI energies relative to the $\mathrm{S}_1$ energy at the FC point are in parentheses.}
	\label{tabactive}
	\begin{ruledtabular}
		\begin{tabular}{lccc}
         & $E(\mathrm{FC},\mathrm{S}_1)$ & \multicolumn{1}{c}{$E(\mathrm{CI})$, 11--12CI} & \multicolumn{1}{c}{$E(\mathrm{CI})$, 13--14CI}\\\hline 
 \multicolumn{4}{c}{SA-CASSCF} \\
(8\textit{e},8\textit{o}) & 3.71                       & 2.18 ($-1.53$) & 2.25 ($-1.46$) \\
(10\textit{e},10\textit{o}) & 2.83                       & 2.01 ($-0.82$) & 1.72 ($-1.12$) \\
(12\textit{e},12\textit{o}) & 3.36                       & 2.58 ($-0.78$) & 2.20 ($-1.16$)  \\[5pt]
          \multicolumn{4}{c}{XMS-CASPT2} \\
(8\textit{e},8\textit{o})   & 2.40                       & 2.17 ($-0.23$) & 2.76 ($+0.36$)  \\
(10\textit{e},10\textit{o}) & 2.39                       & 1.87 ($-0.52$) & 2.71 ($+0.32$)  \\
(12\textit{e},12\textit{o}) & 2.46                       & 1.89 ($-0.57$) & 2.86 ($+0.40$) \\
		\end{tabular}
	\end{ruledtabular}
\end{table}

The computational cost of multireference calculations drastically increases with the increasing size of the active space.
Therefore, for computational efficiency, it is of practical interest to employ as small active spaces as possible when simulating photodynamics.
However, the use of such smaller active spaces has to be carefully benchmarked, since it could severely deteriorate the accuracy.
In this section, we have tested two smaller active spaces for RPSB6, (8\textit{e},8\textit{o}) and (10\textit{e},10\textit{o}),
and compared with the results obtained from the full valence $\pi$ active space [i.e., (12\textit{e},12\textit{o})].

Table~\ref{tabactive} compiles the energies [$E(\mathrm{FC},\mathrm{S}_1)$, $E(\mathrm{CI})$] relative to $E(\mathrm{FC},\mathrm{S}_0)$ obtained by SA-CASSCF and XMS-CASPT2 using these active spaces.
The main features in (12\textit{e},12\textit{o}) CASPT2 calculations, i.e., $E(\mathrm{CI})$ at the 13--14CI being higher than $E(\mathrm{FC},\mathrm{S}_1)$, are reproduced by the calculations with smaller active spaces.
Using XMS-CASPT2, the deviations in $E(\mathrm{FC},\mathrm{S}_1)$ and $E(\mathrm{CI})$ from the full $\pi$-valence results are smaller than 0.10 eV and 0.30 eV, respectively, using both active spaces.
To the contrary, the SA-CASSCF results are very sensitive to the selection of the active spaces, with 0.82 eV error in the worst case.
This encouraging behavior is not surprising because XMS-CASPT2 partially accounts for electronic correlation outside the active space.

For RPSB6, a calculation of XMS-CASPT2 nuclear gradients and derivative couplings with the (8\textit{e},8\textit{o}) active space is about 3 times faster than that with full valence $\pi$ space.
Our results show that the trends in the MECI energies are qualitatively reproduced, implying that a smaller active space may be used in future dynamics simulations using XMS-CASPT2.
To use smaller active spaces in dynamics simulations, however, further validations are warranted to ensure the energy conservation throughout the dynamics. 

\subsection{Basis Sets}

\begin{table}[t]
	\centering
	\caption{Dependence on the size of basis functions. The $\mathrm{S}_1$ energies at the FC point and the MECIs are shown in eV relative to the $\mathrm{S}_0$ energy at the FC point. The MECI energies relative to the $\mathrm{S}_1$ energy at the FC point are in parentheses.}
	\label{tabbasis}
	\begin{ruledtabular}
\begin{tabular}{ldcc}
	            & \multicolumn{1}{c}{$E(\mathrm{FC},\mathrm{S}_1)$} & \multicolumn{1}{c}{$E(\mathrm{CI})$, 11--12CI} & \multicolumn{1}{c}{$E(\mathrm{CI})$, 13--14CI} \\ \hline
    \multicolumn{4}{c}{RPSB3}\\
    cc-pVDZ                  & 3.87 & 2.93 ($-0.93$)                  & 2.56 ($-1.31$)  \\
    aug-cc-pVDZ              & 3.78 & 2.86 ($-0.92$)                  & 2.59 ($-1.20$)  \\
    cc-pVTZ                  & 3.84 & 2.87 ($-0.98$)                  & 2.69 ($-1.15$)  \\
    cc-pVQZ                  & 3.83 & 2.84 ($-0.99$)                 & 2.74 ($-1.09$) \\[5pt]
    \multicolumn{4}{c}{RPSB4}\\
	cc-pVDZ                  & 3.32 & 2.18 ($-1.14$)                 & 2.77 ($-0.56$) \\
	aug-cc-pVDZ              & 3.26 & 2.14 ($-1.12$)                 & 2.81 ($-0.45$)  \\
	cc-pVTZ                  & 3.30 & 2.18 ($-1.13$)                 & 2.93 ($-0.37$)  \\[5pt]
    \multicolumn{4}{c}{RPSB5}\\
	cc-pVDZ                  & 2.78 & 2.02 ($-0.76$)                 & 2.80 ($+0.01$)  \\
	aug-cc-pVDZ              & 2.74 & 2.05 ($-0.68$)                 &  2.81 ($+0.08$)    \\
	cc-pVTZ                  & 2.77 & 2.12 ($-0.65$)                 &  2.89 ($+0.12$)
\end{tabular}
    \end{ruledtabular}
\end{table}

To investigate how the impact of dynamical correlation changes with the size of basis functions,
we have performed molecular geometry optimizations of RPSB3, RPSB4, and RPSB5 with larger basis sets.
We used the aug-cc-pVDZ, cc-pVTZ, and cc-pVQZ basis sets for RPSB3 and the aug-cc-pVDZ and cc-pVTZ basis sets for RPSB4 and RPSB5.
The results are compiled in Table~\ref{tabbasis}.
As the basis-set size becomes larger,
the vertical excitation energies computed by XMS-CASPT2 at the all-\textit{trans} point, $E(\mathrm{FC},\mathrm{S}_1)$, relative to $E(\mathrm{FC},\mathrm{S}_0)$ becomes smaller.
They decrease by 0.01--0.02 eV (cc-pVTZ), 0.04 eV (cc-pVQZ) and 0.05--0.08 eV (aug-cc-pVDZ) compared to that computed using cc-pVDZ.

The MECI energies at the 13--14CI [relative to $E(\mathrm{FC},\mathrm{S}_0)$] become higher when larger basis sets are used, though the changes are small (about $0.1$ eV).
The trend is ascribed to the fact that the differential dynamical correlation effects are more pronounced with larger basis sets,
which can describe more dynamical electron correlation contributions.
Based on these systematic results, we expect that the same holds for RPSB6, for which we did not perform the calculations with larger basis sets.
It is important to note that our results justify the use of small (double-$\zeta$) basis sets in the previous sections and photodynamics simulations in the future. 

\subsection{State Averaging Schemes\label{SA2sec}}

\begin{table}[t]
	\caption{Dependence on the state averaging schemes. The energies at the MECIs are shown in eV relative to the $\mathrm{S}_0$ energy at the FC point. The MECI energies relative to the $\mathrm{S}_1$ energy at the FC point are in parentheses.
	\label{tabsa}}
    \begin{ruledtabular}
	\begin{tabular}{clllllll}
		      & \multicolumn{2}{c}{SA-CASSCF} & \multicolumn{2}{c}{XMS-CASPT2} \\
		      &  \multicolumn{1}{c}{11--12CI}    & \multicolumn{1}{c}{13--14CI}    & \multicolumn{1}{c}{11--12CI}        & \multicolumn{1}{c}{13--14CI}        \\ \hline 
        \multicolumn{5}{c}{SA2}\\
		RPSB3   & 3.46 ($-1.22$)       & 2.51 ($-2.17$)    &  2.93 ($-0.68$)       & 2.31 ($-1.31$)      \\
		RPSB4     & 2.66 ($-1.35$)       & 2.41 ($-1.60$)   &  2.26 ($-0.96$)       & 2.44 ($-0.78$)     \\
		RPSB5       & 2.20 ($-1.78$)       & 2.37 ($-1.01$)   &  1.79 ($-0.92$)       & 2.55 ($-0.16$)   \\
        RPSB6     & 2.58 ($-0.68$)\footnotemark[1] & 2.39 ($-0.89$) & 1.98 ($-0.36$) & 2.63 ($+0.29$) \\[5pt]
        \multicolumn{5}{c}{SA3}\\
		RPSB3       & 3.98 ($-0.81$)       & 2.75 ($-2.04$)  & 2.93 ($-0.93$)       & 2.56 ($-1.31$)    \\
		RPSB4     & 3.08 ($-1.05$)       & 2.52 ($-1.62$)  & 2.18 ($-1.14$)       &   2.77 ($-0.56$)    \\
		RPSB5     &    2.50 ($-0.99$)     & 2.27 ($-1.22$)     &  2.02 ($-0.76$)       &  2.80 ($+0.01$) \\
        RPSB6     &  2.58 ($-0.78$) & 2.20 ($-1.16$) & 1.89 ($-0.58$) & 2.86 ($+0.39$) \\
	\end{tabular}
    \end{ruledtabular}
    \footnotetext[1]{Slightly looser threshold was used due to slow convergence.}
\end{table}

There are two commonly used schemes for state averaging in SA-CASSCF and XMS-CASPT2 when studying the RPSB photodynamics: 
that including the three lowest singlet states\cite{Andruniow2004PNAS,Gozem2012JCTC,Garavelli1998JACS,MunozLosa2011JCTC,Cembran2004JACS,Ruckenbauer2010JPCA,Manathunga2017JPCL,Liu2016JPCB,Frutos2007PNAS,Dokukina2017PhotochemPhotobiol,Luk2015PNAS}
and that including the two lowest states.\cite{Manathunga2016JCTC,Ben-Nun2002PNAS,Gozem2012JCTC,Gozem2012Science,Valsson2010JCTC,Valsson2012JPCL,Page2003JCC,Fantacci2004JPCA,Zhou2014JCC,Szymczak2008JCTC,Mori2010JCP,Schapiro2011JACS}
We have performed geometry optimizations of all the models using both schemes.
For the sake of brevity, we will abbreviate these schemes as SA3 and SA2 schemes.
In addition, we will denote the SA-CASSCF and XMS-CASPT2 calculations with these schemes as SA$n$-CASSCF and SA$n$-XMS-CASPT2, where $n$ is 3 or 2.

Unlike the results obtained by the SA3 scheme, the correlation energy difference between the states at the 11--12 MECI of RPSB5 using SA2-CASSCF is below 0.01 eV.
Consequently, the 11--12CI geometry from SA2-CASSCF closely resembles the SA2-XMS-CASPT2 geometry [root-mean-square deviations between the SA-CASSCF and XMS-CASPT2 geometries are 0.04~\AA~(SA2) and 0.23~\AA~(SA3) for RPSB5].
At the 13--14CI obtained by SA2-CASSCF, the difference between the correlation energy contributions for these two states is about 0.54 eV, which is roughly the half of the difference computed using the SA3 scheme.
Though smaller than that obtained by the SA3 scheme, this differential dynamical correlation is sufficient to shift the 13--14CI geometry toward the FC point along the BLA coordinate [root-mean-square deviations are 0.63~\AA~(SA2) and 0.72~\AA~(SA3) for RPSB5].

The reason why the SA2 correlation energy difference is smaller than that obtained by the SA3 scheme can be explained as follows.
The $\mathrm{S}_0$ ($1\mathrm{A}_g$) and $\mathrm{S}_2$ ($2\mathrm{A}_g$) states have similar electronic characters.\cite{Gozem2012JCTC,Gozem2017CR,Manathunga2017JPCL}
In the XMS-CASPT2 calculations, the state-averaged Fock operator is used to define the zeroth-order Hamiltonian, using which the perturbative correction to the energy is computed.
Using the SA2 scheme is equivalent to excluding one $\mathrm{A}_g$ state from the state-averaged Fock operator in the SA3 scheme, and therefore,
the PT2 correction to $\mathrm{A}_g$ and $\mathrm{B}_u$ will be larger and smaller, respectively, than the correction obtained based on the SA3 scheme.
With SA2, therefore, the differences between the PT2 energy for the $\mathrm{B}_u$ and $\mathrm{A}_g$ states become smaller than with SA3.

Despite such differences between the SA3 and SA2 schemes,
the trends in the MECI energies with respect to the chromophore lengths are similar in XMS-CASPT2,
and the MECI energies with both schemes agree within 0.4~eV (Table \ref{tabsa}).
The MECI geometries with both schemes are also similar when XMS-CASPT2 is used: the root-mean-square deviations between them are 0.11~\AA~and~0.12~\AA~for the 11--12CI and 13--14CI of RPSB5, respectively.
From this observation, it is unclear to us which scheme is superior to the other in the XMS-CASPT2 calculations.
Nevertheless, there will be certain situations that one scheme is preferred over another to obtain more accurate results.
For example, a recent molecular dynamics study\cite{Manathunga2017JPCL} has shown that the mixing between $\mathrm{S}_2$ and $\mathrm{S}_1$ becomes more important in the polar environment,
especially near the FC point,
because the $\mathrm{A}_g$ states are more stabilized due to the environment than the $\mathrm{B}_u$ state.
This implies that it is essential to include $\mathrm{S}_2$ in the multistate calculations (i.e., use the SA3 scheme) when solvent or protein environments are considered in the simulation.

\section{Conclusion}

In this study, we have optimized the equilibrium and MECI geometries of the RPSB using many structural and computational models.
The largest model, RPSB6, consisted of 54 atoms, for which we used the (12\textit{e},12\textit{o}) active space.
We found that, as one increases the model size, the XMS-CASPT2 energies at the MECIs (relative to the all-\textit{trans} minimum) become lower at the 11--12CI and higher at the 13--14CI, respectively.
These results are in sharp contrast with those obtained by SA-CASSCF.
The differences between the MECI structures obtained by XMS-CASPT2 and SA-CASSCF are ascribed to the fact that the diradicaloid ($1\mathrm{A}_g$) and charge transfer ($1\mathrm{B}_u$) states are stabilized by dynamical correlation to different degrees.
As a result, the MECI structures are shifted toward the FC point along the BLA coordinate.
Various active spaces, basis sets, and state averaging schemes are tested.
It is found that the XMS-CASPT2 results are less sensitive to the selection of active spaces than the SA-CASSCF results.
Our results suggest that the basis-set dependence is small (though larger basis sets do increase the differential dynamical correlation effect), justifying the use of
double-$\zeta$ basis sets when simulating the photodynamics of these chromophores.
The differential dynamical correlation effects are also shown to be more pronounced
when the S$_2$ ($2\mathrm{A}_g$) state is included in the SA-CASSCF and XMS-CASPT2 calculations.
These results should serve as the basis for choosing the computational models used in the future studies on RPSB photodynamics.

\section{Acknowledgments}
This article is part of the special issue in honor of the 80th birthday of Professor Michael Baer.
We are grateful to the Air Force Office of Scientific Research Young Investigator Program (Grant No.~FA9550-15-1-0031) for generous financial support.
The development of the program infrastructure has been in part supported by National Science Foundation [ACI-1550481 (JWP) and CHE-1351598 (TS)].

\end{document}